\documentclass[preprint2]{aastex}
\usepackage{color}

\newcommand{\Mo}{\it M_{\odot}}
\shorttitle{Propeller-Effect Interpretation of 4U~1608$-$52 and Aql~X-1}
\shortauthors{Asai et al.}

\begin{document}


\title{Propeller-Effect Interpretation of MAXI/GSC Light Curves
of 4U~1608$-$52 and Aql~X-1 and application to XTE~J1701$-$462}


\author{
K. Asai\altaffilmark{1}, M. Matsuoka\altaffilmark{1}, 
T. Mihara\altaffilmark{1}, M. Sugizaki\altaffilmark{1}, 
M. Serino\altaffilmark{1}, S. Nakahira\altaffilmark{2},
H. Negoro\altaffilmark{3}, Y. Ueda\altaffilmark{4},
and K. Yamaoka\altaffilmark{5}
}
\email{kazumi@crab.riken.jp}

\altaffiltext{1}{MAXI team, RIKEN, 2-1 Hirosawa, Wako, Saitama 351-0198, Japan}
\altaffiltext{2}{ISS Science Project Office, ISAS, JAXA, 2-1-1 Sengen,
Tsukuba, Ibaraki 305-8505, Japan}
\altaffiltext{3}{Department of Physics, Nihon University,
1-8-14 Kanda-Surugadai, Chiyoda-ku, Tokyo 101-8308, Japan}
\altaffiltext{4}{Department of Astronomy, Kyoto University,
Kitashirakawa, Oiwake-cho, Sakyo-ku, Kyoto 606-8502, Japan}
\altaffiltext{5}{Institute of Space and Astronautical Science,
JAXA, 3-1-1 Yoshino-dai, Chuo-ku, Sagamihara, Kanagawa 252-5210, Japan}


\begin{abstract}
We present the luminosity dwell-time distributions during the hard
states of low-mass X-ray binaries containing a neutron star,
4U~1608$-$52 and Aql~X-1, observed with MAXI/GSC\@.  The luminosity
distributions show a steep cut-off in the low-luminosity side at
$\sim1.0 \times 10^{36}$ erg s$^{-1}$ in both the two sources.  The
cut-off implies a rapid luminosity decrease in their outburst decay
phases, and the feature can be interpreted as due the propeller
effect.  We estimated the
surface magnetic field of the neutron star
to be (0.5--1.6) $\times 10^8$~G in 4U~1608$-$52 and (0.6--1.9)
$\times 10^8$~G in Aql~X-1 from the cut-off luminosity.  We applied
the same propeller mechanism to the similar rapid luminosity decrease
observed in the transient Z-source, XTE~J1701$-$462, with RXTE/ASM.
Assuming that spin period of the neutron star is in the order of
milliseconds, the observed cut-off luminosity deduces surface magnetic
field in the order of $10^9$~G.
\end{abstract}


\keywords{Stars: neutron --- X-rays: binaries  --- X-rays: individual
(Aql~X-1, 4U~1608$-$52, XTE~J1701$-$462) --- X-rays: propeller effect}

\section{Introduction}

Low-mass X-ray binaries with a neutron star (NS-LMXB) consists of an
old weakly-magnetized neutron star (NS) ($<10^{10}$~G) and an evolved
late-type companion star.  According to their timing properties and
spectral variations represented by color--color and
hardness--intensity diagrams, NS-LMXBs are classified into two groups:
Z sources and Atoll sources (e.g., Hasinger \& van der Klis 1989).
However, what causes the difference between the Z sources and the
Atoll source, is still under debate.

A theoretical study of the NS magnetic field evolution due to the
accretion suggested that the magnetic field of Z sources
($\sim10^9$~G) are greater than those of Atoll sources ($\sim10^8$~G)
(Zhang \& Kojima 2006).  This agrees with the following observational
results.  The magnetic field of a Z source, Cyg~X-2, was estimated to
be $2.2 \times 10^9$~G from the observed horizontal-branch
oscillations and the beat frequency model (Focke 1996).  Those of
Atoll sources, 4U~1608$-$52 and Aql~X-1, were estimated to be
(1.4--1.8) $\times 10^8$~G and 1 $\times 10^8$~G, respectively from
the presumable propeller effect (Chen et al. 2006; Zhang et al. 1998a;
Campana et al. 1998).
\footnote{ Their values were calculated assuming a distance of 3.6~kpc
  for 4U~1608$-$52 and that of 2.5~kpc for Aql~X-1.  In this study, we
  employ distances of 4.1~kpc and 5.0~kpc for 4U~1608$-$52 and for
  Aql~X-1, respectively.  If these values were applied in their study,
  their magnetic field values would be (1.6--2.0) $\times 10^8$~G for
  4U~1608$-$52 and 2 $\times 10^8$~G for Aql~X-1. } 
Regarding 4U~1608$-$52, Weng and Zhang (2011) also derived an estimate
of $\sim 10^8$~G from the interaction between the magnetosphere and
the accretion flow.  On the other hand, the magnetic-field strength of a
transient Z source XTE~J1701$-$462, which exhibited the transition
from the Z-type to the Atoll-type in the color--color diagram during
the outburst decay phase (Homan et al. 2010), was estimated to be
$\sim$(1--3) $\times 10^9$~G from the interaction between the
magnetosphere and the radiation-pressure-dominated accretion disk
(Ding et al. 2011).  However, Titarchuk, Bradshaw, and Wood
(2001) suggested that both Z sources and Atoll sources have a very low
surface magnetic fields of $\sim10^6$--$10^7$~G based on their
magneto-acoustic wave model and the observed kHz QPOs.

As for the two Atoll sources, 4U~1608$-$52 and Aql~X-1, further fine
state-transition behaviors were investigated.  Wachter et al. (2002)
and Maitra and Bailyn (2008) identified three distinct states:
Outburst, extended Low-Intensity state (LIS), and True Quiescence (TQ)
from the correlation between X-ray and optical or IR data.  Matsuoka
and Asai (2013), hereafter referred to as MA2013, proposed four
states; Soft, Hard-High, Hard-Low, and No-accretion (recycled pulsar)
state according to the mass-flow rate and the NS magnetic-field
strength.  The Soft and the Hard-High states are characterize by the
accretion disk states, which are optically thick (Soft) or thin
(Hard-High).  The two Hard states, Hard-High and Hard-Low, are
classified in terms of the propeller effect.  As the mass accretion
onto the NS decreases, its magnetosphere expands, and then the
accretion flow is finally restricted by the centrifugal barrier in the
Hard-Low state.

Although a number of observational evidences indicating state
transitions featured by a simultaneous flux and spectral change have
been obtained so far, these interpretations on the underlying physical
process are still rather confused.  Chen, Zhang, and Ding (2006) and
Zhang, Yu, and Zhang (1998) proposed the propeller effect for an 
interpretation on observed flux decreases accompanied with
soft-to-hard state transitions.  However, Maccarone and Coppi (2003)
pointed out that the propeller effect is not a sole cause for all the
observed state transitions; they reported that the luminosity in the
Hard-to-Soft transition in the rising phase is greater than that in
the Soft-to-Hard transition in the decay phase by a factor of $\sim$5
or more.  If the propeller effect is the sole cause, both the
transitions in the rise and the decay may occur at the same
luminosity.  
MA2013 proposed the four-state picture including both the
inner disk transition (Soft to Hard-High,Abramowicz et al. 1995)
and the propeller effect
(Hard-High to Hard-Low) to explain the behaviors of 4U~1608$-$52 and
Aql~X-1, as mentioned above.

In this paper, we present luminosity dwell-time distributions of
4U~1608$-$52, Aql~X-1, and XTE~J1701$-$462, and propose the consistent
interpretation of these profiles base on the propeller effect with
their intrinsic magnetic fields.  This also makes observationally
develop the simplified picture of various NS-LMXB states proposed by
MA2013.  In section~2, we describe the MAXI/GSC observations of
4U~1608$-$52 and Aql~X-1, and the data analysis.  There, we report the
rapid luminosity decreases in the outburst decay phases and determine
the cut-off luminosities from the luminosity dwell-time distributions.
In section~3, we estimate the magnetic-field strengths from the
cut-off luminosities.  We discuss the validity of the propeller effect
interpretation base on the obtained parameters.  Subsequently, we
apply the same method to XTE~J1701$-$462 RXTE/ASM light curve, and
summarize the results in section~4.

\section{Observation and Analysis}

The GSC (Gas Slit Camera: Mihara et al. 2011; Sugizaki et al. 2011a)
on the MAXI (Monitor of All-sky X-ray Image: Matsuoka et al. 2009)
payload detected two outbursts from 4U~1608$-$52 and three outbursts from
Aql~X-1 from August 2009 to September 2012
(MJD = 55058--56180)
(Morii 2010, Sugizaki et al. 2011 for 4U~1608$-$52 and 
Yamaoka et al. 2011 for Aql~X-1).
We used the GSC 2--10~keV light-curve data on
the public archive~\footnote{$<$http://maxi.riken.jp/$>$.}
provided by the MAXI team.
We also utilized the 15--50~keV light-curve data provided by
Swift (Gehrels et al. 2004)/ BAT (Burst Alert Telescope: Barthelmy et al. 2005)
team.~\footnote{
$<$http://heasarc.gsfc.nasa.gov/docs/swift/results/transients/$>$.}
The GSC count rates are converted to the luminosities
by assuming that the spectrum is Crab-like (Kirsch et al. 2005) and
employing the source distances of $4.1\pm0.4$~kpc in 4U~1608$-$52 and 
$5.0\pm0.9$~kpc in Aql~X-1 (Galloway et al. 2008).

Figure~1 shows GSC light curves, BAT light curves, and the BAT/GSC
hardness ratios for 4U~1608$-$52 and Aql~X-1.
We identified the spectral state (Soft state or Hard states)
from the BAT/GSC hardness ratio using the same method in Asai et al. (2012).
The mark ``S'' in figure~1 indicates the Soft state period which is clearly
recognized by the hardness ratio of BAT/GSC.
A rapid decrease of GSC luminosity in the 2--10~keV band is seen at the
transition time from the Soft state to Hard state.
This rapid luminosity decrease and spectral hardening occurred
by the inner disk transition proposed by MA2013.
The roman numerals in the figures denote the Hard-state periods.
These Hard-state periods can be divided into two sub-states: one with
a luminosity at around $\sim10^{36}$ erg s$^{-1}$ and another with
that below the detection limit ($\sim 3\times 10^{35}$ erg s$^{-1}$).
The two sub-states correspond to the ``Hard-High'' and the
``Hard-Low'' states, respectively, defined in MA2013.
They are also considered to coincide with the LIS and the TQ in
Wachter et al. (2002), respectively, from the levels of X-ray
intensities in the sub-states.
These sub-state periods are summarized in table~1.

While the Hard-High states are clearly recognized in the three hard
states, 4U~1608$-$52 (I), 4U~1608$-$52 (III) and Aql~X-1(II), 
they are hard to see in the other hard states, that is,
4U~1608$-$52 (II), Aql~X-1 (I), Aql~X-1 (III), and Aql~X-1 (IV). 
In the latter cases, the source changed immediately
from the Soft state to the Hard-Low state; thus we
cannot discriminate the level of the Hard-High to Hard-Low transition
at which the propeller effect occurred.
To investigate the propeller effect hereafter, we focus on the former
three periods, 4U~1608$-$52 (I), 4U~1608$-$52 (III) and Aql~X-1(II).

In figure~2, light curves of the selected three Hard-state periods are
magnified.  All the three curves show a rapid decrease when the
luminosity decreased below the threshold of $\sim10^{36}$ erg
s$^{-1}$.  We call the threshold luminosity starting the rapid
decrease ``cut-off luminosity''.  In order to evaluate the cut-off
luminosity, we create luminosity dwell-time distributions during the
three Hard-state periods in figure~3.  In the Appendix, the luminosity
distributions for typical light-curve profiles are presented.

In the period 4U~1608$-$52 (I), the luminosity distribution has a peak
at $0.015 \times10^{38}$ erg s$^{-1}$ (figure~3a), which corresponds
to the flux plateau in the Hard-High state (figure~2a).  In the
luminosity below the peak, the dwell-time is very small.  This implies
that the luminosity decreased rapidly below the Hard-High plateau.
Therefore, the cut-off luminosity is $1.0\times10^{36}$ erg s$^{-1}$
from the luminosity distribution.
In the period 4U~1608$-$52 (III), rapid luminosity decrease occurred
several times at several luminosity levels (figure~2b).
This variation of cut-off luminosity is also seen in the dwell-time
distribution in figure~3b,
in which the cut-off luminosity is ranging in 
(0.75--1.0)$\times10^{36}$ erg s$^{-1}$.
We adopted the common value for the cut-off luminosity of 4U~1608$-$52
to be 1.0$\times10^{36}$ erg s$^{-1}$.
In the period Aql~X-1 (II), the rapid luminosity decrease
is clearly seen.  The cut-off luminosity is
estimated from the lower edge of the peak in the histogram to be
$1.3\times10^{36}$ erg s$^{-1}$ (figure~3c).  
These cut-off luminosities are also indicated in figure~2
by dashed lines.

These rapid luminosity decreases occurred in the Hard-state periods
(from Hard-High to Hard-Low),
and thus differ from the state transition due to the 
inner disk transition 
(from Soft to Hard-High).  
Namely, the rapid luminosity decrease occurred at the luminosity lower than
the transition luminosity of the Soft-to-Hard-high.

\section{Discussion and Application}

\subsection{Surface Magnetic Fields Derived from Luminosity Dwell-Time Distributions}

We extracted luminosity dwell-time distribution during the Hard-state
period including both the Hard-High and Hard-Low, and determined the
cut-off luminosity below which the luminosity start to decrease
rapidly.  This cut-off luminosity corresponds to the transition
luminosity from the Hard-High to the Hard-Low, when the propeller
effect become effective (MA2013).  If the cut-off
luminosity is due to the propeller effect, we can derive the surface
magnetic field on the NS ($B$) using the following equation
(MA2013):~\footnote{ Here, an orthogonal dipole magnetic field is
  assumed and the relations $P_{\rm mag}=B^2/(4\pi)$, $\mu = BR^3$ are
  used, following Ghosh and Lamb (1979).  }
\begin{eqnarray}
B \ = \ 2.6 \times 10^7 \eta^{-7/4} \ 
\biggl(\frac{P}{{\rm 1\ ms}}\biggr)^{7/6}  
\biggl(\frac{L}{10^{36}\ {\rm erg~s^{-1}}}\biggr)^{1/2}
\nonumber
\end{eqnarray}
\begin{eqnarray}
\times
\biggl(\frac{M}{1.4\,\Mo}\biggr)^{1/3}\ \biggr(\frac{R_{\rm ns}}{10^6~{\rm cm}}\biggl)^{-5/2} {\rm G}
\end{eqnarray}
where $P$, $M$, $R_{\rm ns}$ denote the spin period, mass, and radius of
the NS, respectively. 
The term $L$ denotes the luminosity at which the propeller effect
occurs and the co-rotation
radius equals the Alfv\'{e}n radius.  
The model dependence factor $\eta \sim$ 0.5--1 is obtained from
the definition of the Alfv\'{e}n radius $R_{\rm A} = \ \eta\ R_{\rm A0}$.
The ideal Alfv\'{e}n radius $R_{\rm A0}$ is defined in equation (27)
in Ghosh and Lamb (1979).

Consequently, we can derive the surface magnetic field of the NS
to be (0.5--1.6)$\times 10^8$~G for 4U~1608$-$52 and
(0.6--1.9)$\times 10^8$~G for Aql~X-1.
Here, we adopted spin periods of 1.62~ms for 4U~1608$-$52
(Hartman et al. 2003) 
and 1.82~ms for Aql~X-1 (Zhang et al. 1998b).

\subsection {Comparison with Previous understanding as Propeller Effect}

As the evidence of the propeller effects, Soft-to-Hard spectral
transitions accompanied with a sudden flux decrease during a outburst
decay phase observed in Aql~X-1 (Campana et al. 1998; Zhang et
al. 1998) as well as 4U~1608$-$52 (Chen, Zhang \& Ding 2006) have
ever been suggested.
However, such an incident that characterized by the simultaneous flux
decrease and spectral change may occur not only by the propeller effect
but also in the
inner disk-state transition from the optical thick to the
thin one (e.g., Asai et al. 2012).  Actually, Maccarone and Coppi
(2003) pointed out that the propeller effect is not the sole cause of
the spectral transitions because the transition luminosities at the
outburst rise and the decay phases are significantly different.
Therefore, to identify the propeller effect correctly, we need to
distinguish it from the inner disk transition.
The previous studies on the propeller effects did not take account of
the disk transition properly.
Thus, their estimates on
the propeller level may not be appropriate.

Using the MAXI/GSC and Swift/BAT data covering the wide energy band
with more frequent observations
and moderate sensitivity, we were able to
separate the two transitions clearly,
although we cannot see a hardness change of BAT/GSC hardness ratio
by propeller effect clearly since the data is poor in statistics.
This is remarkably different
from the previous works. However, our estimates of the magnetic field
strengths on both 4U~1608$-$52 and Aql~X-1 were almost consistent with
the previous results (although ours are slightly small). 
This is
because the two transition luminosities for the
inner disk-state transition
and
the propeller effect are rather close (within a factor of $\sim$ 5).

\subsection {Knee Features in the Light Curve of the Outburst Decay}
\label{knee}
Powell, Haswell, and Falanga (2007), hereafter referred to as PHF2007,
reported a knee feature (they labeled it as a ``brink'') in the light
curve of outburst decay phase, due to the change of the accretion rate
coupled with the disk irradiation.  If the ``brink'' occurred, the
decay tendency would change from the exponential to the linear.  This
feature has been recognized in both NS-LMXBs and BH-LMXBs.
In fact, our light curves (figure 1) also show the ``brink'' feature
characterized by the transition from the exponential to the linear
decline in the Soft state of 4U~1608$-$52 marked with S$_{\rm III}$,
and that of Aql~X-1 marked with S$_{\rm IV}$.
The luminosity at ``brink'' is $\sim$1.6$\times10^{37}$ erg s$^{-1}$
in 4U~1608$-$52 and $\sim$3.8$\times10^{37}$ erg s$^{-1}$ in Aql~X-1.
Campana et al. (2013) reported the ``brink'' features in both S$_{\rm II}$
and S$_{\rm III}$ of Aql~X-1.

The rapid luminosity decrease in the Hard-High to the Hard-Low state
transition discussed so far may resemble the ``brink'' feature.
However, the number of data points are too small
to make detailed analysis, and consequently,  
we can fit the decay curve with either of a single linear function,
or a single exponential function.
Note that 4U~1608$-$52 (I) has 6 data points in the decay part, and
Aql~X-1 (II) has only 2 data points.
The decay part of 4U~1608$-$52 (III) is difficult to be defined
because the rapid luminosity decrease occurred several times.
Therefore, it was difficult for us to perform useful fittings.
However, the rapid luminosity decrease occurred at the end of the long-lasting
($\sim$ 100 days) Hard-high state.
It does not occur when the flux is decaying monotonically from the outburst
peak in which the ``brink'' was observed.
Thus, the feature of transition from the Hard-High to the Hard-Low state
is most probably the propeller effect.
Here, note that we derived the cut-off luminosity of the propeller 
effect making the luminosity dwell-time distribution in figure~3
because of no useful fitting analysis.

\subsection{Application to XTE~J1701$-$462}

The propeller effect occurred in the Hard-state period in 4U~1608$-$52
and Aql~X-1.  However, the inner
disk state and propeller effect are
independent issues. 
Depending on the magnetic fields and spin period,
the propeller effect can occur in the higher or the lower luminosity
than that of the inner
disk transition.
A rapid luminosity decrease was
observed in a transient Z source, XTE~J1701$-$462 (Lin et al. 2009a;
Fridriksson et al. 2010).  The root cause of that rapid decrease has
not been understood yet.  In this subsection, we apply our
understanding of the propeller effect to the observed XTE~J1701$-$462
data.

\subsubsection{Previous works on XTE~J1701$-$462}

The transient X-ray source on the Galactic plane, XTE~J1701$-$462,
flared up to the super-Eddington luminosity in 2006, and then
continued the activity for more than 18 months.  The RXTE (Bradt et
al. 1993) / ASM (Levine et al. 1996) continuously monitored the flux
is down to near-quiescence.  The source transformed from the Z-type
into the Atoll-type as the luminosity decreased (Homan et al. 2007a,
2007b, 2010).  It also exhibited a rapid luminosity decrease in the
outburst decay phase.  We plot, in figure 4a, the ASM light curve
obtained from the data archive provided by the RXTE-ASM team at MIT
and NASA/GSFC~\footnote{ $<$http://xte.mit.edu/$>$.}, where observed
count rates are converted to the luminosities in 2--10~keV band by
assuming that the spectrum is Crab-like (Kirsch et al. 2005) and the
distance is 8.8~kpc (Lin et al. 2009b).  The rapid decrease started on
MJD$\sim$54303, which accords to the epoch when the transition in the
color--color diagram from the Z-type to the Atoll-type occurred (Lin
et al. 2009a).  The curves before the onset, MJD$\sim$54300, can be
fitted with a Gaussian function of a width $\sim$70~d, as overlaid on
the data in figure 4a.  

Lin et al. (2009a) also reported that the Atoll-state period can be
divided into the Soft state (SS) and the Hard state (HS).
The transition from the Soft to the Hard state occurred on MJD$\sim$54312 (July
31, 2007).
%, which is significantly after the onset of the rapid
%decrease.  
This means that the source remained in the Soft state when the rapid
decrease started.  We consider it possibly due to
the propeller effect, and then estimate the surface magnetic field
on the NS using the same methods applied to 4U~1608$-$52
and Aql~X-1.

%luminosity
%dwell-time distributions to XTE~J1701$-$462.

\subsubsection{Luminosity dwell-time distribution}

We derived the luminosity dwell-time distributions in figure~4b for
the light curve of figure~4b, where the expected distribution for the
Gaussian decline model in the light curve is overlaid.
The rapid luminosity decrease after MJD = 54300 corresponds to the
lower cut-off in the luminosity distributions at the level pointed by
the arrow, which is estimated to be $1.8\times10^{37}$ erg s$^{-1}$.
The cut-off level is also shown on the light curve in figure~4a.

\subsubsection{Cut-off luminosity and magnetic fields}

The cut-off luminosity coincides with that when the transition from
the Z source to the Atoll source occurred (Lin et al. 2009a).  This
cut-off is not likely to be a ``brink'' by PHF2007 (described in
subsection 3.2) since the decrease is exponential (not linear) [see
  figure~2 in Fridriksson et al. (2010)].  Thus, it can be interpreted
as due to the propeller effect where the Z--Atoll state transition may
occur as a consequence of the mass-accretion decrease.  The kHz QPOs
were observed from XTE~J1701$-$462 in both the Z-state and the
Atoll-state (Sanna et al. 2010).  Based on the obtained QPO variation,
they proposed that the accretion flow around the NS changed in the
Z--Atoll state transition.  This hypothesis is consistent with our
propeller-effect interpretation.

Since XTE~J1701$-$462 had kHz QPOs, let us assume that the spin period of the
NS is in the order of milliseconds.  Then the surface magnetic field
strength is deduced to be the order of $10^9$~G.  Ding et al. (2011)
derived the surface magnetic field strength of XTE~J1701$-$462 as
$\sim$(1--3) $\times 10^9$~G from the interaction between the
magnetosphere and the radiation-pressure-dominated accretion disk.
Our result is consistent with their result.

Fridriksson et al. (2010) discussed a possibility of the propeller
effect for this rapid decrease, and mentioned that there are ``several
serious problems in general interpretation'' of the outburst decay
rate in NS transitions.  One of the ``serious problems'' is in the
fact that not only NS-LMXB but also black-hole transients have shown 
similar rapid-decreasing decay features (Chen et al. 1997:Jonker et al. 2004).
Indeed, the rapid-decreasing decay features were observed in 
NS-LMXB transients, black-hole transients and also Cataclysmic variables 
(e.g., Gilfanov et al. 1998 and references therein).
The common interpretation of ``outer-disk thermal instability''
(e.g.Meyer \& Meyer-Hofmeister 1981; Mineshige \& Wheeler 1989)
is proposed
(Chen et al. 1997; King \& Ritter 1998; Gilfanov et al. 1998).
The ``brink" in PHF2007 is due to the ``outer-disk thermal instability'' and
we were able to distinguish them from the propeller effect. 
Therefore, we consider that there is no ``serious problem''
in the propeller effect interpretation on the observed rapid
luminosity decrease in this NS-LMXB.

\subsubsection{Modification of the simplified model of LMXB}

We have seen in the previous subsection that the rapid luminosity
decrease of XTE J1701$-$462 can be understood as the propeller effect.
The simplified picture of NS-LMXBs in MA2013 proposed the
classification of the spectral states of 4U~1608$-$52 and Aql~X-1 by
the combination of the inner
disk-state and the propeller-state (table~2).
Let us define the transition luminosity of the disk-state and the
propeller-state as $L_{\rm disk}$ and $L_{\rm prop}$, respectively.
In the case of $L_{\rm prop} < L_{\rm disk}$ including 4U~1608$-$52
and Aql~X-1 (top in table~2), the observed X-ray spectrum is supposed
to undergo the Soft, Hard-High, Hard-Low, and no-accretion state in
this order as the luminosity decreases.  In the other case of $L_{\rm
  disk} < L_{\rm prop}$ like XTE~J1701$-$462 (bottom in table~2), it
will change in the order of the Z-like Soft, Atoll-like Soft (Atoll
banana), and Atoll-like Hard (Atoll island) state, instead.

%Since a NS-LMXB has the Alfv\'{e}n radius depending on the luminosity
%and the co-rotation radius proper to the NS, the propeller effect is
%expected to occur at some luminosity 

Depending on the magnetic-field strength and the spin period which are
specific parameters for each NS-LMXB, the propeller effect may occur
in either the Soft state or the Hard state.
%
%%%% This is not a topic here. (MS)
%Thus, to determine the propeller effect, we
%have to distinguish it from other similar incidents such as the disk
%transition, the brink by PHF2007.  
%
%
When a propeller effect occurs, it is expected that the thermal
emission from the NS surface would decrease and the non-thermal one
from the surrounding gas would become conspicuous.  In addition,
Rappaport et al. (2004) noticed that even in the propeller effect the
spectrum will change in a weak intensity, since a part of inflow gases
would leak onto the NS.
There is a 1~Hz QPO observations in another transient LMXB, SAX~J1808.4$-$3658
(Patruno et al. 2009). The timing information might be an important information
in connection with  the propeller effect.
For further investigation of the propeller
effect, various observations and analysis of light curves and spectra
for many sources would be needed.

\section{Summary}

We analyzed the light curves and the luminosity dwell-time
distributions in the Hard-state periods of 4U~1608$-$52 and Aql~X-1,
obtained by MAXI/GSC\@.  The luminosity distributions show a peak with
a steep cut-off in the lower luminosity side.  This cut-off
corresponds to a rapid luminosity decrease in the outburst decay
phase.  The cut-off luminosity is $\sim1.0 \times 10^{36}$ erg
s$^{-1}$ in both 4U~1608$-$52 and Aql~X-1.  

Any NS-LMXB has two qualitative radii; the Alfv\'{e}n radius and the
co-rotation radius.  When the former exceeds the latter as the
luminosity decreases, the propeller effect is expected to occur.  The
cut-off can be interpreted as due to the effect.
The obtained cut-off luminosities imply the
surface magnetic fields on the NS, (0.5--1.6)
$\times 10^8$~G in 4U~1608$-$52 and (0.6--1.9) $\times 10^8$~G in
Aql~X-1.

We applied the propeller-effect interpretation to the light curve of a
transient Z-source, XTE~J1701$-$462, obtained by RXTE/ASM.  This
source also showed a rapid luminosity decrease, but that occurred in
the Soft state.  The deduced surface magnetic field is in the order of
$10^9$~G and consistent with the previous work.  Thus, the rapid
luminosity decrease in this source can also be due to the propeller
effect regardless of the spectral states.

\acknowledgments

The authors acknowledge the MAXI team for MAXI operation
and for observing and analyzing real-time data.  
They also thank the Swift-BAT team for providing the excellent quality data
in the public domain.
This research was partially supported by the Ministry of Education, 
Culture, Sports, Science and Technology (MEXT), 
Grant-in-Aid No. 20244015 and 24340041.

\appendix

\section{Luminosity Dwell-Time Distribution}

In order to study the relation between the light curve and its
luminosity dwell-time distribution, we present, in figure~5, those for
three typical light curves: a random variation
in the logarithmic scale,
an exponential decrease, and a Gaussian decrease.  
In case of a random variation in the logarithmic scale,
the dwell-times are constant and independent of
luminosity values if the histogram-bins are chosen to have an equal width
in the logarithmic scale  (top panel).
When the luminosity varies between $L_{\rm min}$
and $L_{\rm max}$, the luminosity dwell-time distributions exhibit
equal heights between $L_{\rm min}$ and $L_{\rm max}$.
In the exponential decrease, the dwell-time also becomes constant
because the luminosity decrease is represented by a straight line in
the logarithmic scale (middle panel).
In the case of a Gaussian decrease, luminosity decreases more rapidly,
which reduces the dwell-time in in lower luminosities
(bottom panel).

\clearpage

\begin{figure*}
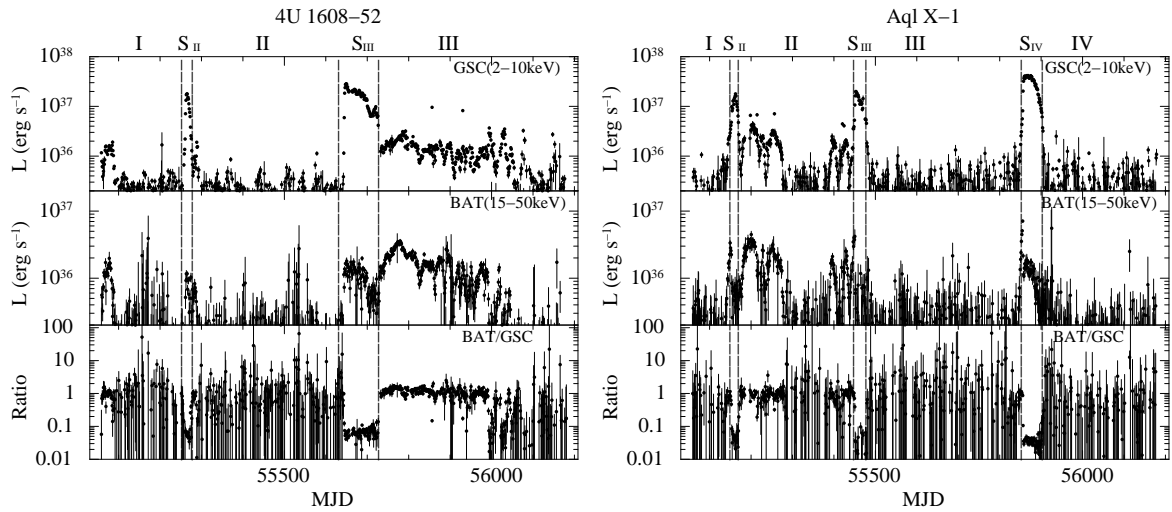

\plotone{fig1-a.eps}
\plotone{fig1-b.eps}
\caption{
GSC light curves (2--10~keV:top), BAT light curves (15--50~keV:middle),
and the hardness ratios (BAT/GSC:bottom)
for 4U~1608$-$52(left) and Aql~X-1(right)
since the start of the MAXI observation in August 2009 (MJD = 55058).
Vertical error-bars represent 1-$\sigma$ statistical uncertainties.
Roman numerals denote the Hard-state
periods and ``S$_{\rm II-IV}$'' indicates the Soft-state periods.  
}
\label{fig1}
\end{figure*}

\clearpage

\begin{figure}
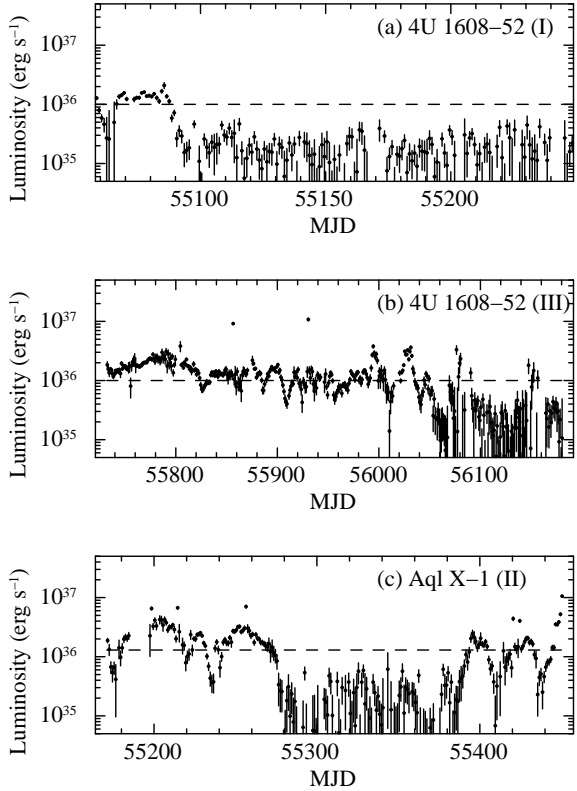

    \plotone{fig2-a.eps}
    \plotone{fig2-b.eps}
    \plotone{fig2-c.eps}
\caption{ Magnified light curves of three Hard-state periods of 
  4U~1608$-$52 (I), (III) and Aql~X-1 (III) in figure~1.
  Dashed lines represent cut-off luminosities 
  indicated by arrows in figure~3.  The two
  jumping points in 4U~1608--52 (III) at MJD = 55856 and 55930, and
  the five jumping points in Aql~X-1 (II) at MJD = 55198, 55214,
  55256, 55420, and 55424 are due to type-I X-ray burst.  }
\label{fig2}
\end{figure}

\begin{figure*}
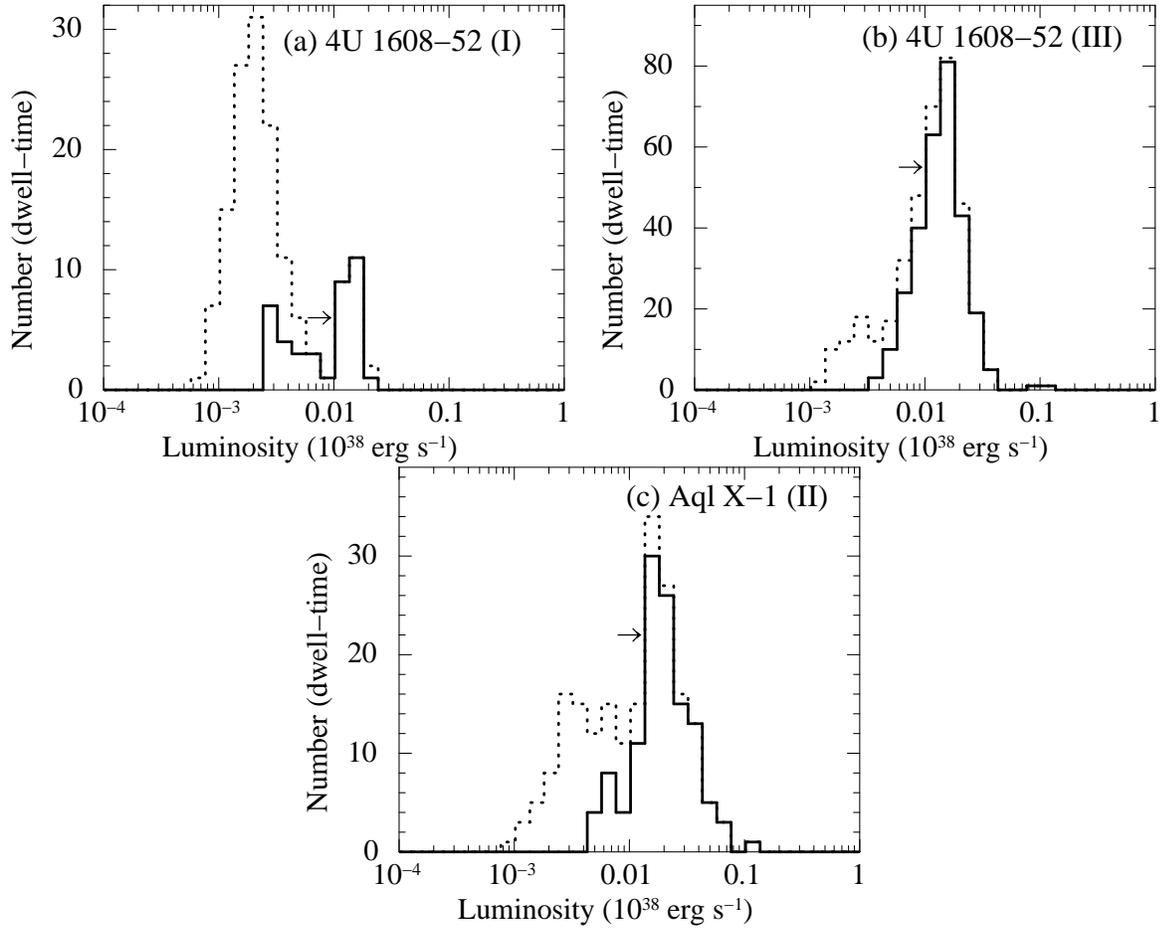

%\epsscale{.70}
    \plotone{fig3-a.eps}
    \plotone{fig3-b.eps}
    \plotone{fig3-c.eps}
\caption{
Luminosity dwell-time distributions in three Hard-state periods of
(a) 4U~1608$-$52 (I), (b) 4U~1608$-$52 (III),
and (c) Aql~X-1 (II).
The width of each bin is chosen to be equal in the logarithmic scale. 
The histogram with the dotted line includes
data with a significance of more than $1\sigma$,
and that with the solid line employs
data with a significance of more than $4\sigma$.
The arrow in each figure indicates a steep cut-off in the lower side,
which corresponds to a rapid luminosity decrease.
This cut-off luminosity is shown by the dashed lines in figure~2.
}
\label{fig3}
\end{figure*}

\begin{figure*}
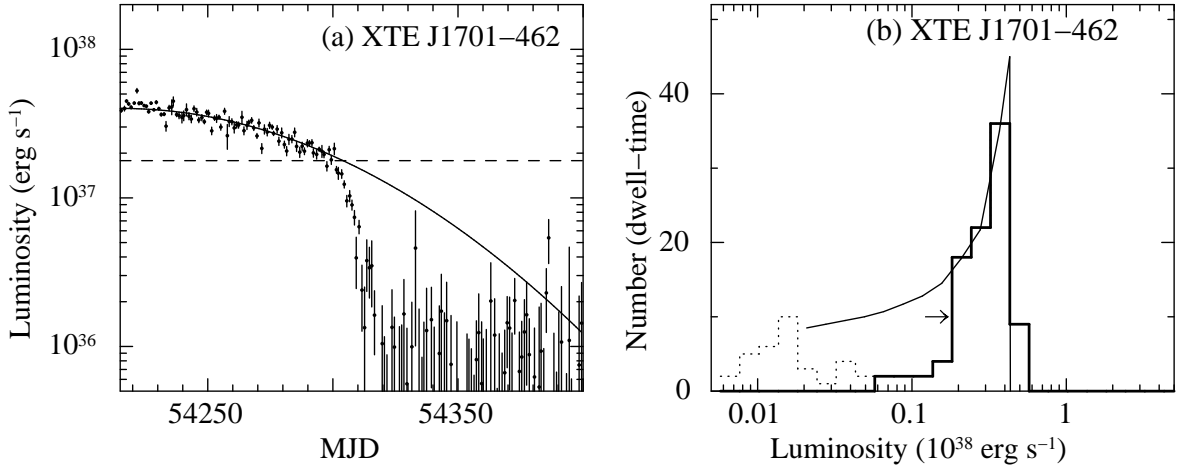

%\epsscale{.70}
    \plotone{fig4-a.eps}
    \plotone{fig4-b.eps}
  \caption{
Light curve (a) and luminosity dwell-time distribution (b) of XTE~J1701$-$462
from MJD 54215 to 54400.
In the light curve (a), the data until MJD$=$54300 are fitted
with a Gaussian function, as drawn in a solid curve.
In the luminosity distribution (b), 
the histogram with the dotted line includes
data with a significance of more than $1\sigma$,
and that with the solid line employs
data with a significance of more than $4\sigma$.
The luminosity distribution for the model is drawn in a solid curve.
The arrow indicates a steep cut-off in the lower side,
which corresponds to the rapid luminosity decrease.
This level of the cut-off luminosity is indicated with a dashed line on
the light curve.
}
\label{fig4}
\end{figure*}

\begin{figure*}
%\epsscale{1.10}
  \plotone{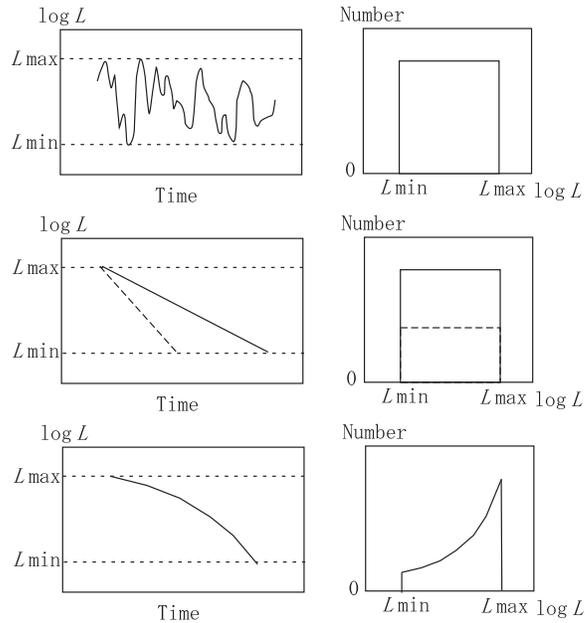}
  \caption{
Schematic drawing of light curves (left panels) and luminosity
dwell-time distributions (right panels) in three typical cases of
random variation in the logarithmic scale (top), an exponential
decrease (middle), and a Gaussian decrease (bottom).  }
\label{fig5}
\end{figure*}

\begin{table*}
\caption{Summary of Hard-state periods of 4U~1608$-$52 and Aql~X-1.}
\label{tab1}
\begin{center}
\begin{tabular}{cccc}
\tableline\tableline
Period & Start date\tablenotemark{a} (MJD) & End date\tablenotemark{a} (MJD) &
Sub-state in Hard state \tablenotemark{b} \\
\tableline
\multicolumn{4}{c}{4U~1608$-$52} \\
\tableline
I  & 55058 & 55253 & Hard-High and Hard-Low\\
II & 55274 & 55643 & Hard-Low \\ 
III \tablenotemark{c}
 & 55732 & 56180 &  Hard-High and Hard-Low\\ 
\tableline
\multicolumn{4}{c}{Aql~X-1} \\
\tableline
I  & 55058 & 55152 & Hard-Low \\ 
II & 55171 & 55450 &  Hard-High and Hard-Low\\
III & 55479 & 55855 & Hard-Low \\
IV & 55904 & 56178 & Hard-Low \\ 
\tableline
\end{tabular}
\tablenotetext{a}
{
Soft-to-Hard transitions occurred on the start dates
and Hard-to-Soft transitions occurred on the end dates.
}
\tablenotetext{b}
{Defined by MA2013. 
%Most of the Hard-Low were below the detection limit
%of MAXI/GSC because of the quiescent period of the Hard-Low,
%although the transition from Hard-High to Hard-Low was detected by MAXI/GSC.  
See text.}
\tablenotetext{c}
{During this period, ``mini'' outbursts are seen at around
MJD = 55994, 56031, 56080, and 56152.
The hardness ratio variation behaves as Hard-to-Soft transitions 
in the first two (around MJD = 55994 and 56031),
although the luminosities
did not exceed the typical transition luminosity
of $1\times10^{37}$ erg s$^{-1}$. 
The latter two (around MJD = 56080 and 56152) were
not clear because of poor statistics of the BAT data.}
\end{center}
\end{table*}

\begin{table*}
%\caption{States and propeller effect of three LMXBs.}
\caption{Spectral states and propeller effect in the three NS-LMXBs:
  4U~1608$-$52, Aql~X-1, and XTE~J1701$-$462}
\label{tab2}
\begin{center}
\begin{tabular}{clcccccc}
\tableline\tableline
Luminosity & State & \multicolumn{4}{l}{Propeller} & \multicolumn{2}{c}{Disk} \\
\tableline
\multicolumn{8}{l}{4U~1608$-$52 and Aql~X-1} \\
\tableline
high           & Soft    & no,   & $R_{\rm A}$ $<$ &  $R_{\rm c}$ &
& optically & thick \\
$\updownarrow$ & Hard-High & no, & $R_{\rm A}$  $<$ &  $R_{\rm c}$ &
&           & thin \\
low            & Hard-Low & yes, &            & $R_{\rm c}$ & $<$ $R_{\rm A}$
&           & thin \\
\tableline
\multicolumn{8}{l}{XTE~J1701$-$462} \\
\tableline
high           & Soft (Z-like Soft)     & no, & $R_{\rm A}$ $<$ & $R_{\rm c}$ &
& optically &thick \\
$\updownarrow$ & Soft (Atoll-like Soft) & yes, & & $R_{\rm c}$ & $<$ $R_{\rm A}$
& & thick \\
low            & Hard-High (Atoll-like Hard) & yes, & & $R_{\rm c}$ & $<$ $R_{\rm A}$ & & thin \\
\tableline
\end{tabular}
\end{center}
\end{table*}

\end{document}